\documentclass[a4 paper, superscriptaddress, 11pt]{article}

\usepackage[utf8]{inputenc}
\usepackage[T1]{fontenc}
\usepackage{amsmath, amsfonts, amssymb}
\usepackage{graphicx}
\usepackage{fancyhdr}
\usepackage[left=3cm, right=3cm, top=3cm, bottom=3cm]{geometry}
\usepackage{enumitem}
\usepackage{multirow}
\usepackage[numbers,sort&compress]{natbib}% Citation support using natbib.sty

\begin{document}

\begin{center}
\textbf{\Large{Factorization, Supersymmetry, Coherent States and Classical Trajectories}}
\end{center}

\begin{center}
\textrm{\normalsize{\textbf{$^\ast$Musongela\  Lubo, $^\ast$Kikunga\  Kasenda\ Ivan, $^\dagger$Likwolo\  Katamba\ Stanislas}}}\\
\end{center}

\begin{center}
{\textit{\scriptsize{$^\star$Physics Department, Faculty of Sciences, University of Kinshasa, P.O. Box 190 Kin\ XI, Kinshasa, D.R.Congo}}}\\
\textit{\scriptsize{$^\dagger$Faculty of Agronomy, University of Uele, P.O. Box 670 Isiro, Isiro,
D.R.Congo}}\\
\end{center}

\begin{center}
{\scriptsize{Email: musongela.lubo@unikin.ac.cd, ivan.kikunga@unikin.ac.cd, likwolostanislas@yahoo.fr}}\\
\end{center}
\

\begin{abstract}
A generalization of coherent states has been developed in the context of supersymmetric quantum mechanics. For many cases, no link has been made with the corresponding classical system. 
In this work, we consider simple superpotentials and compare the classical trajectories and the mean values of the position operator in these states. The mean value of the position operator can be written as a power series in time with coefficients whose relations to the superpotential are given in the general case.\\ These coefficients imply integrals and recursion formulas.
The method used reproduces exactly what is known for the harmonic oscillator. It is extended to study a family of systems which encompasses the harmonic oscillator. We also consider a third degree superpotential. The time scale after which the mean value of the position operator and the classical trajectory begin differing significantly is evaluated.\\

\begin{center}
\textbf{Résumé}\\
\end{center}

Une généralisation des états cohérents a été développée dans le contexte de la mécanique quantique supersymétrique. Dans de nombreux cas, aucun lien n'a été établi avec le système classique correspondant. Dans ce travail, nous considérons des superpotentiels simples et comparons les trajectoires classiques et les valeurs moyennes de l'opérateur position dans ces états. La valeur moyenne de l'opérateur position peut être écrite comme une série entière du temps avec des coefficients dont les relations au superpotentiel sont données dans le cas général.\\
Ces coefficients impliquent des intégrales et des formules de récurrence. La méthode utilisée reproduit exactement ce qui est connu pour l'oscillateur harmonique. Elle est ensuite étendue pour étudier une famille de systèmes qui englobe l'oscillateur harmonique. Nous considérons également un superpotentiel du troisième degré. L'échelle de temps après laquelle la valeur moyenne de l'opérateur de position et la trajectoire classique commencent à différer de manière significative est évaluée.

\end{abstract}


\section{Introduction}
\

%%%%%%%%%%%%%%%%%%%%%%%%%%%%%%%%%%%%%%%%%%%%%%%%%%%%%%%%%%%%%%%%%%%%%%%%%%%%%%%%%%%%%%%%%%%%%%%%%%%%%%

This work is essentially devoted to the link between the classical and the "quantum"\footnote{We mean by this, the mean value of the position operator.} trajectories of two systems described by the same Lagrangian. We focus here on specific models which will be described later.\\

With a given Lagrangian, the classical trajectory is a solution of the Euler-Lagrange equations. The Heisenberg uncertainty relation, at the quantum level, forbids a perfect localisation in position. This leads to the notion of "quantum" trajectory given by the mean value of the position operator. It has to be considered for all quantum states permitted by the system under consideration. The Ehrenfest theorem basically states that classical and "quantum" trajectories coincide for a narrow family of potentials and quantum states.\\

Hence the interest of the standard coherent states, which reconcile classical and "quantum" trajectories for the harmonic oscillator. They inspired important developpments in mathematical and theoretical physics as well as works linked to technology\\

Among the works concentrainting on mathematical aspects, one can cite \cite{Art01-01, Art01-02,Art01-03,Art01-04,Art01-05,Art01-06,Art01-07, Art01-08,Art01-09,Art01-10, Art01-11,Art01-12,Art01-13,Art01-14,Art01-15,Art01-16,Art01-17,Art01-18,Art01-19, Art01-20,Art01-21,Art01-22,Art01-23,Art01-24,Art01-25,Art01-26,Art01-27,Art01-28, Art01-29,Art01-30,Art01-31,Art01-32,Art01-33,Art01-34,Art01-35}. These studies mostly concentrate on the quantum operators involved and their representations on specific Hilbert spaces. The irreducibility and the unitarity of these representations were studied. Symmetry (ordinary and super) arguments play a very important role in these works, as well as the dimension of the Hilbert spaces involved.
In this context, different generalizations of coherent states were proposed. This implies one has to be careful when using the word "coherent" since it does not always have the same meaning for different authors.\\

Concerning Theoretical Physics, many works use coherent states to study light beams\cite{Art01-36,Art01-37,Art01-38,Art01-39,Art01-40,Art01-41,Art01-42}. Other concentrate on molecular or atomic systems, especially the Morse potential which is known to describe the motion of the nuclei of diatomic molecules in a quite simple and effective way \cite{Art01-43,Art01-44,Art01-45,Art01-46,Art01-47,Art01-48,Art01-49,Art01-50}. One can also find articles related  to coherent states for other systems. This is for example the case of the Poschl-Teller potential\cite{Art01-51,Art01-52,Art01-53}, the Coulomb problem \cite{Art01-54,Art01-55}, the anharmonic oscillators\cite{Art01-56} or the infinite potential well\cite{Art01-57}.\\

There are many other fields in which coherent states are used. To be brief, let us mention their possible applications to domains such as cryptography and quantum teleportation\cite{Art01-58,Art01-59,Art01-60,Art01-61,Art01-62}, cosmology\cite{Art01-63} or particle physics\cite{Art01-64}.\\

Concerning Supersymmetric quantum mechanics(SUSYQM) and coherent states, the literature is very rich, so that any attempt about citations is necessarily incomplete. Here, we shall simply cite \cite{Art01-01,Art01-02,Art01-04,Art01-11,Art01-23,Art01-25,Art01-27,Art01-28,Art01-29,Art01-31,Art01-65,Art01-66,Art01-67,Art01-68}.\\

In this work, we study the behaviours of the classical trajectories and the mean value of the position operator in particular coherent states. Other approaches exist which analyse quantum and classical probabilities. They have been treated for example by R.W.Robinett \cite{Art01-69} and CC.Real et al \cite{Art01-70}. 

Some of the computations performed in this paper require mathematical formula which can be found in \cite{Art01-71}.\\

%%%%%%%%%%%%%%%%%%%%%%%%%%%%%%%%%%%%%%%%%%%%%%%%%%%%%%%%%%%%%%%%%%%%%%%%%%%%%%%%%%%%%%%%%%%%%%%%%%%%%

The path taken here is the following. For the harmonic oscillator, the superpotential is a one degree polynomial. To get an insight into  the subject, we study  an ad hoc system whose superpotential is  polynomial. For simplicity and to ensure normalizability, we take it to be of third degree. The physical potential is then a sixth degree polynomial whose classical trajectories can be computed. 
The coherent states can be obtained in closed form. We then proceed to the calculation of the mean value of the position operator for these states. Comparison is then made with the classical trajectories.\\

The basic result of this paper is that the mean value of the position operator can be written as a power series in time with coefficients whose relations to the superpotential are given in the general case. These coefficients imply integrals and recursion formulas.\\

The paper is organized as follows. The second section is a quick reminder of the formalism of Supersymmetry (SUSY) and coherent states that we need. The third section deals with the calculation of the mean value of the position operator in general. We apply it to a simple toy model in the fourth section. The treatment of the harmonic oscillator is put in the Appendix A and shows that the analysis captures what is known by other methods. Appendix B deals with a family of superpotentials related to the harmonic oscillator.

\section{A Quick Reminder of Coherent States and SUSYQM}
\

Different inequivalent generalizations of coherent states have been proposed in the literature\cite{Art01-01,Art01-05,Art01-06,Art01-17,Art01-19,Art01-22,Art01-26,Art01-44,Art01-54}. They all rely on some of the properties exhibited in the case of the harmonic oscillator\cite{Art01-05,Art01-16,Art01-19}. Among other things, these states saturate the Heisenberg inequality, are eigenstates of the annihilation operator and realise a decomposition of the identity operator. Moreover, and that is of physical significance, the mean value of the position operator in these states reproduces exactly the classical trajectories (for the harmonic oscillator). 

Let us now turn to SUSYQM. The context we work in was proposed by Molski \cite{Art01-01}. Consider a system described by a potential $V_0$ which admits a ground state energy. With the ground state wave function and its energy, one can construct a new potential $V_1$ whose spectrum and eigenstates are related to the one corresponding to $V_0$. The Hamiltonians $H_0$ and $H_1$ can be factorized with the same operators appearing in reverse order\cite{Art01-31}. These operators lead to a graded algebra\cite{Art01-01,Art01-02,Art01-04,Art01-11,Art01-23,Art01-25,Art01-27,Art01-28,Art01-29,Art01-31,Art01-65,Art01-66,Art01-67,Art01-68}.\\

We are studying, in this context, a one dimensional system with a rescaled undimensional  coordinate q. Suppose one can write the physical potential $V(q)$ of such a system in terms of a function $x(q)$  by the relation
\begin{equation}
\label{eq01}
V(q)-E_0=\frac{1}{2}\Big\lbrack x^2(q)+\frac{dx(q)}{dq} \Big\rbrack,
\end{equation}
$E_0$ being the ground state energy. It is physically not essential since only differences in energies can be measured.\\
Then one can factorize the Hamiltonian
\begin{equation}
\label{eq02}
\hat{H}=\hat{A^\dag}\hat{A}+E_0,
\end{equation}
where the operators $\hat{A}$ and $\hat{A^\dag}$ have the form
\begin{equation}
\label{eq03}
\hat{A}=\frac{1}{\sqrt{2}} \Big\lbrack \frac{d}{dq}-x(q) \Big\rbrack  ,\  \hat{A^\dag}=\frac{1}{\sqrt{2}}\Big\lbrack -\frac{d}{dq}-x(q) \Big\rbrack,
\end{equation}
so that
\begin{equation}
\label{eq03a}
\Big\lbrack \hat{A} , \hat{A^\dag} \Big\rbrack = - \frac{dx(q)}{dq}.
\end{equation}
The function $x(q)$ is called the superpotential. The equations \eqref{eq03} and \eqref{eq03a} are reminiscent of the harmonic oscillator.(Note that $\hat{A}$ and $\hat{A^\dag}$ are not necessairly ladder operators in the general case). Using that similarity, the coherent states are defined by Molski\cite{Art01-01} as the eigenfunctions of the operator $\hat{A}$. \\
It should  be noted that although defined as eigenstates of a well motivated operator, the states proposed by Molski also saturate the bound of  the Heisenberg uncertainty relation for some well defined generalized  coordinates.
Using \eqref{eq03} one is led to an equation of separable variables. The solution, for the coherent states as understood in this context reads
\begin{equation}
\label{eq04}
\psi_H(q, \alpha)=N \exp{\Big\lbrack \sqrt{2}\alpha q+\int_0^q  x(\xi) \, d\xi \Big\rbrack},
\end{equation} with N a normalization constant.
In \eqref{eq04}, $\alpha$ is a complex variable which characterizes the coherent states.
When trying to implement the SUSY treatment to a system whose physical potential is known, the difficult part is the resolution of the Riccati equation given in \eqref{eq01}.\\
For the harmonic oscillator, one introduces the rescaled variables
\begin{equation}
\label{eq05}
Q = \sqrt{\frac{\hbar}{m\omega}}\  q,  \  P =\sqrt{m \omega\hbar} \ p
\end{equation}
and the rescaled potential
\begin{equation}
\label{eq06}
U (Q)=\hbar\omega \ V (q).
\end{equation}
The superpotential is given by
\begin{equation}
\label{eq07}
x(q)=-q,
\end{equation}
so that the coherent state takes the form
\begin{equation}
\label{eq08}
\psi_H(q, \alpha)=N \exp{( \sqrt{2} \alpha q-\frac{1}{2}q^2 )}.
\end{equation}\\

One should keep in mind that factorizing a Hamiltonian does not automatically lead to ladder operators. Moreover the operators involved in the factorization do not always form a closed algebra. This important issue was adressed by Mielnik, Fernandez, Rosas Ortiz,... (See \cite{Art01-27,Art01-28,Art01-29,Art01-65,Art01-66,Art01-67}). These  papers are among the first to study generalized coherent states for supersymmetric partners of the harmonic oscillators.\\

In this work, we do not propose a new definition for coherent states. We take the one  proposed by Molski\cite{Art01-01} and try to figure out the mean value of the position operator and compare it to the classiclal trajectory in some cases.\\

One of the most important results concerning classical and "quantum" trajectories is the Ehrenfest theorem. It states that the mean values of the position and momentum operators obey the following differential equations
\begin{eqnarray}
\begin{split}
\label{art01}
\frac{d}{dt}\langle \hat{x} \rangle(t)&=\frac{1}{m} \langle \hat{p}\rangle(t),\\
\frac{d}{dt}\langle \hat{p} \rangle(t)&=- \langle V'(\hat{x}) \rangle.
\end{split}
\end{eqnarray}
These expressions are "close" to the Hamilton equations for the classical trajectory whose position and momentum are denoted by $x_{cl}$ and $p_{cl}$:
\begin{eqnarray}
\begin{split}
\label{art02}
\frac{d}{dt}x_{cl}(t)&=\frac{1}{m} p_{cl}(t),\\
\frac{d}{dt}p_{cl}(t)&=- V'(x_{cl}(t)).
\end{split}
\end{eqnarray}
The difference quantum mechanics brings  resides in the derivative of the potential. One has to consider its mean value for the state under consideration. It is a common understanding that for potentials which are not quadratic(with the possibilty of a linear term), one has in general 
\begin{equation}
\label{art03}
\langle V'(\hat{x}) \rangle (t) \not= ( V' (\langle \hat{x} \rangle )) (t).
\end{equation}
This says that in most cases, the classical and "quantum" trajectories will not coincide. Thus, our question concerning the time scale at which a significant departure occurs is justified. One has nevertheless to keep in mind for a torough analysis that this departure for early times may be compensated if the quantum trajectory is comprized in a band around the classical one.

\section{Our Approach}
\

It is well known that Quantum Mechanics can be expressed in two different pictures which in fact are equivalent. In the Schrodinger picture, the physical operators do not depend on time while the wave function does. The Heisenberg picture does the opposite. The two points of vue are related by the evolution operator. The quantity we are interested in is the mean value(written here in the Schrodinger picture):
\begin{equation}
\label{eq12}
q_{\text{moy}}(t) =\frac{\langle \psi_{S ,t} \lvert \hat{q}\rvert \psi_{S ,t} \rangle }{\langle \psi_{S, t} \lvert \psi_{S, t} \rangle }.
\end{equation}
One has to realize that the states given in \eqref{eq04} were obtained in the Heisenberg picture: the wave function does not depend on time. To pass to the Schr\"{o}dinger picture one has to use the evolution operator
\begin{equation}
\label{eq13}
\lvert \psi_{S,t} \rangle = \exp {( -\frac{i}{\hbar}\hat{H}t )} \rvert \psi_H\rangle,
\end{equation}
where the Hamiltonian takes the form  
\begin{equation}
\label{eq14}
\hat{H}=\hbar \omega \Big\lbrack \frac{\hat{p}^2}{2} + \hat{V}(q) \Big\rbrack.
\end{equation}
The factorization of the Hamiltonian and the evolution operator can be used to obtain the wave function in the Schr\"{o}dinger picture as a power series
\begin{equation}
\label{eq15}
\lvert \psi_{S,t}\rangle = \exp(- \frac{i}{\hbar} E_0 t ) .\sum_{k=0}^{n} (-i)^k \frac{(\omega t)^k}{k!}(\hat{A^\dagger}\hat{A})^k \rvert \psi_H \rangle.
\end{equation}
The mean value then reads
\begin{equation}
\label{eq16}
\langle \psi_{S ,t} \lvert \hat{q} \rvert \psi_{S ,t}\rangle = \sum_{m=0}^{\infty} \sum_{n=0}^{\infty} \frac{(-1)^m i^{m+n}}{m! n!} t^{m+n}  \langle \psi_H \lvert \hat{H}^n \hat{q} \hat{H}^m \rvert \psi_H \rangle.
\end{equation}
The relations \eqref{eq15} and \eqref{eq16} show that we need to calculate powers of the product $\hat{A^\dag}\hat{A}$. For this we need a few ingredients. The state, designated by $\vert \psi_H \rangle$, is an eigenstate of the operator $\hat{A}$;
\begin{equation}
\label{eqbb01}
\hat{A} \vert \psi_H \rangle = \alpha \vert \psi_H \rangle.
\end{equation}
The following relation, 
\begin{equation}
\label{eq17}
\hat{A^\dag}=-\hat{A}-\sqrt{2}x(q)
\end{equation}
is obtained by adding the two relations given in \eqref{eq03}. One thus has 
\begin{equation}
\label{eqbb02}
\hat{A}^{\dag} \hat{A} = - \left( \hat{A}+\sqrt{2}x(q) \right)\hat{A},
\end{equation}
so,
\begin{equation}
\label{eq18}
\hat{A^\dag}\hat{A}\lvert \psi_H \rangle = \alpha \Big\lbrack -\alpha-\sqrt{2}x(q)\Big\rbrack \rvert \psi_H \rangle.
\end{equation}
To evaluate $\left(\hat{A}^\dag \hat{A}\right)^2$, one will need to compute the following quantities:
\begin{eqnarray}
\begin{split}
\label{eqcc01}
\hat{A} \left( x(q)\vert \psi_H \rangle \right)&= \frac{1}{\sqrt{2}} \Big\lbrack \ \frac{dx(q)}{dq} + \sqrt{2}\alpha x(q) \ \Big\rbrack \vert \psi_H \rangle,  \\
\hat{A}\left(\frac{dx(q)}{dq}\vert \psi_H \rangle\right)&= \frac{1}{\sqrt{2}} \ \Big\lbrack \frac{d^2 x(q)}{dq^2} + \sqrt{2}\alpha \frac{dx(q)}{dq} \ \Big\rbrack \vert \psi_H \rangle.
\end{split}
\end{eqnarray}
The following relation is usefull; it is also obtained from the expressions given in \eqref{eq03}.
\begin{equation}
\label{eq19}
\frac{d}{dq} \lvert \psi_H \rangle = \Big\lbrack \sqrt{2}\alpha + x(q) \Big\rbrack \rvert \psi_H \rangle.
\end{equation}
With all these elements, we find :
\begin{eqnarray*}
\label{eqcc02}
&\left(\hat{A^\dag}\hat{A}\right)^2 \vert \psi_H \rangle = \alpha\left(\hat{A}+\sqrt{2}x(q)\right)
\Big\lbrack \hat{A}\left(\alpha +\sqrt{2}x(q)\right)\vert \psi_H \rangle \Big\rbrack, \\
&= \alpha^3 \left(\hat{A}+\sqrt{2} x(q)\right)\vert \psi_H \rangle + \alpha \sqrt{2} \left(\hat{A}+\sqrt{2} x(q)\right)\Big\lbrace \hat{A} \  \Big\lbrack x(q)\vert \psi_H \rangle \Big\rbrack \Big\rbrace.
\end{eqnarray*}
Finally, we obtain the second order contribution to the state 
\begin{equation}
\label{eq20}
(\hat{A^\dag}\hat{A})^2 \lvert \psi_H \rangle = \Big\lbrack  \alpha \frac{\sqrt{2}}{2} \frac{d^2x(q)}{dq^2} + \left( 2\alpha^2+ \sqrt{2} \alpha x(q) \right) \frac{dx(q)}{dq} + \left( \sqrt{2}\alpha x(q)+\alpha^2 \right)^2  \Big\rbrack \rvert \psi_H\rangle.
\end{equation}
This suggests the introduction of special functions $f_n$ such that
\begin{equation}
\label{eq21}
(\hat{A^\dag}\hat{A})^n \lvert \psi_H \rangle = f_n(q,\alpha) \lvert \psi_H \rangle .
\end{equation}
From the previous considerations, one derives the recurrence formula
\begin{equation}
\label{eq22}
f_{n+1}=-\frac{1}{2} \frac{\partial^2f_n}{\partial q^2} - \Big\lbrack \sqrt{2} \ \alpha + x(q) \Big\rbrack \frac{\partial f_n}{\partial q} - \Big\lbrack  \sqrt{2} \alpha x(q) + \alpha^2 \Big\rbrack f_n.
\end{equation}
Note that
\begin{equation}
\label{eq23}
f_0(q,\alpha)=1.
\end{equation}
Working in the position representation, the time dependent wave function 
\begin{equation}
\label{eq24}
\psi_S (q, t, \alpha)=\exp{(-\frac{i}{\hbar} E_0 t)}\sum_{m=0}^{\infty} \frac{(-i\omega t)^m}{m!}\Big\lbrack (\hat{A^\dag}\hat{A})^m \psi_H(q,\alpha)\Big\rbrack
\end{equation}
takes the form
\begin{equation}
\label{eq25}
\psi_S (q, t, \alpha)=\exp{(-\frac{i}{\hbar} E_0 t)}\sum_{m=0}^{\infty} \frac{(-i\omega t)^m}{m!}f_m(q,\alpha)\psi_H(q,\alpha).
\end{equation}
The position mean value is then given by an infinite double sum
\begin{equation}
\label{eq26}
\langle \psi_{S ,t} \lvert \hat{q} \rvert \psi_{S ,t}\rangle = \sum_{m=0}^{\infty} \sum_{n=0}^{\infty} \frac{i^{m+n }(-1)^n}{m!n!} C_{m, n} \ (\omega t)^{m+n},
\end{equation}
where the coefficients $C_{m, n}$ are the integrals
\begin{equation}
\label{eq27}
C_{m, n}= \int_{-\infty}^{+\infty} dq \ q \ f^\star_m (q,\alpha) f_n (q,\alpha) \psi_H^\star (q,\alpha) \psi_H (q,\alpha)\ .
\end{equation}
The double summation can be rewritten as a power series in the time parameter
\begin{equation}
\label{eq28}
q_{moy}(t)= \langle \psi_{S ,t} \lvert \hat{q} \rvert \psi_{S ,t}\rangle = \sum_{l=0}^{+\infty} (\omega t)^l \Omega_l,
\end{equation}
with
\begin{equation}
\label{eq29}
\Omega_l=i^l (-1)^l\sum_{m=0}^{l} \frac{(-1)^m}{m!(l-m)!}C_{m, l-m}.
\end{equation} 
Explicitly, we can write
\begin{eqnarray}
\begin{split}
\label{eqa02}
\Omega_0 &= C_{0, 0},  \\
\Omega_1 &= -i \left( C_{0,1} - C_{1,0} \right), \\
\Omega_2 &= \frac{1}{2!} \left( -C_{0,2} + 2 C_{1,1} - C_{2,0} \right), \\
\Omega_3 &= i \big\lbrack \ \frac{1}{2!} \left( C_{2,1} -  C_{1,2} \right) +  \frac{1}{3!} \left( C_{0,3} - C_{3,0} \right) \ \big\rbrack, \\
\Omega_4 &=  \frac{1}{4!} \left( C_{0,4} +  C_{4,0} \right) -  \frac{1}{3!} \left( C_{1,3} + C_{3,1} \right) + \frac{1}{2! \ 2!} \ C_{2,2},\\
&\cdots
\end{split}
\end{eqnarray}
The equation \eqref{eq28} is written with the fact that the quantity $\langle \psi_{S ,t} \vert \psi_{S ,t}\rangle$ is an unessential constant which is usually fixed to unity.\\
One can consider that \eqref{eq28} and \eqref{eq29} gives the answer to our question about the mean value of the position operator in this context.
For any practical case, one has to compute the integrals of \eqref{eq27} and make the appropriate summation.\\
One has to point out technical difficulties. The first one is that the recursion relations of \eqref{eq22} can quickly lead to large formulas even for simple superpotentials.
The second one is that it is not always possible to have an analytical expression for the integrals appearing in \eqref{eq27}.
Third, one has to be careful about the order at which one can stop in the series given by \eqref{eq28} to obtain a reliable estimate.

To test our approach, we first used it on the harmonic oscillator. The results are good and reproduce some important characteristics of the coherent states as known in the literature \cite{Art01-05}. The solution to the classical equation of motion is given by
\begin{equation}
\label{eqq01}
q_{cl}(t) = C_1 \cos{\omega t}+ C_2 \sin{\omega t},
\end{equation}
where $ C_1$ and $C_2$ are constants. This can be written as a power series
\begin{equation}
\label{eqq02}
q_{cl}(t)=A_0 + A_1 \omega t + A_2 (\omega t)^2 + \cdots
\end{equation}
where one has 
\begin{equation}
\label{eqq03}
\frac{A_2}{A_0}=-\frac{1}{2}, \ \frac{A_3}{A_1}=-\frac{1}{6}, \cdots
\end{equation}
We write the mean value of the position operator in our states using \eqref{eq28}.It can be shown that the same relations are verified i.e.
\begin{equation}
\label{eqqq02}
\frac{\Omega_2}{\Omega_0}=-\frac{1}{2}, \ \frac{\Omega_3}{\Omega_1}=-\frac{1}{6}, \cdots
\end{equation}
To keep our presentation light, we have put this treatment in the Appendix A.

\section{A Toy Model}
\

In this paper, we construct a superpotential such that the corresponding coherent states  normalizability is clear and the computation of the mean values not too complicated.
The first case studied by Molski\cite{Art01-01}, after the harmonic oscillator, was the Morse potential. It would then seem natural  to immediately treat  that case with our approach since we have  derived general expressions for the mean value of the position operator. This study is postponed for the time being for reasons which are explained below.\\
The Morse potential goes to a constant when the distance between  the nuclei is increased, which is at odds with the harmonic oscillator. As a consequence, the harmonic oscillator displays only a discrete  energy spectrum  which contains an infinite number of members, while the Morse potential has a finite number of such states. The Morse potential also contains a continuum spectrum(related to ionisation) which has no counterpart in the harmonic oscillator case. This in itself means one has to be cautious when comparing the two potentials or trying to borrow results from  one case to the other. One should keep in mind that even for Hamiltonians  isospectral to the harmonic one, the limiting process from one to the other is not transparent concerning coherent states (See formula (4.2) in \cite{Art01-66}).\\

The qualitative differences between the harmonic and the Morse systems is the reason for which we are postponing the study of  potentials like those of Wei Hua and Kratzer-Fues which have an important  significance in theoretical  chemistry.\\

This led us to consider a toy model  which is confining like the harmonic oscillator and leads  to  comprehensive  analytical  approximations. It is however clear that  full numerical computations will have to be performed  for the potentials cited just above, which means going above the W.K.B. treatment\footnote{Wentzel-Kramers-Brillouin treatment (approximation).} for example.\\

The generalization of coherent states considered here was introduced in \cite{Art01-01}. The link with the superpotential is given by the formula \eqref{eq04}. It is clear that the wave function given by that formula will not be square integrable for all superpotentials $x(q)$.For example, if the superpotential is  positive everywhere and goes to infinity with increasing distances $q$, the integral appearing in  \eqref{eq04} will also go to infinity. For positive parameters $ \alpha $, this will lead to a wave function which is not normalizable. We did not include here considerations about the behaviour of the potential near the origin. \\

For the harmonic oscillator, the superpotential is a linear function of the position. The next non trivial thing is to study a polynomial superpotential. 
A second degree superpotential, in this scheme, is readily seen to lead to a non normalizable generalized coherent state. This leads us to study a toy model whose superpotential is given by a third degree polynomial:
\begin{equation}
\label{eq30}
x(q) = x_1 q -\frac{1}{3} x_1^2 q^3,
\end{equation}
\begin{equation}
\label{eq31}
\text{with} \ x_1 \leq 0,
\end{equation}
\hspace{50mm}where $x_1$ is a free parameter.\\
The corresponding physical potential is a sixth degree polynomial (See \eqref{eq01} and \eqref{eq06})
\begin{equation}
\label{eq32}
U(Q)= \hbar\omega ( U_0 + U_2 Q^2 + U_4 Q^4 + U_6 Q^6).
\end{equation}
Its coefficients are related to those of the rescaled superpotential by
\begin{equation}
\label{eq33}
U_0 = 0; \ U_2 =  0; \ U_4 = -\left(\frac{\hbar}{m \omega}\right)^2  \frac{x_1^3}{3}; \  U_6 = \left(\frac{\hbar}{m \omega}\right)^3  \frac{x_1^4}{18}.
\end{equation}
It has to be noted that if one begins with the potential, the characteristic frequency is given by 
\begin{equation}
\label{eq34}
\hbar \omega=\frac{U^3_4}{U^2_6}.
\end{equation}

From the equations written above, $x_1$ being negative, formula \eqref{eq34} does not lead to any problem, both sides of the equation being positive. One can deduce the relation $\hbar \omega = - 12 x_1$ where the left and right sides  are again positive. The important point is that the potential is confining, leading to bounded classical trajectories.

This potential has the form given in Fig.1.
\begin{figure}[ht]
\centering
\includegraphics[scale=0.70]{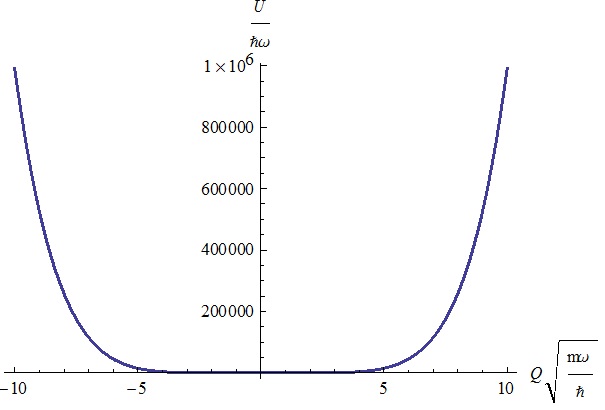}
\caption{{\footnotesize{The Potential for the Toy Model for the values $U_4=1$ and $U_6=1$.}}}
\end{figure}
Classically, all the trajectories are bounded and periodic, with the period
\begin{equation}
\label{eq35}
T_{cl}=4\sqrt{\frac{2}{m}} \int_0^{Q_+} \frac{1}{\sqrt{E-U(Q)}} dQ\
\end{equation}
which depends on the energy of the system because $Q_+$ is such that $E - U(Q_+)=0$.\\
We now analyze the mean value of the position operator for the corresponding coherent states and compare them with the classical trajectories. From \eqref{eq22}, one could easily see that the functions $f_n$ will be polynomials in the variable $q$. This will highly simplify the computations. We thus introduce the coefficients $\tilde{f}_{n,k} (\alpha)$ by
\begin{equation}
\label{eq36}
f_n (q,\alpha)=\sum_{k=0}^{3n} \tilde{f}_{n,k}(\alpha) q^k
\end{equation}
\begin{equation}
\label{eq37}
\text{with} \ \tilde{f}_{0,0}=\tilde{f}_{0,0}^\star=1.
\end{equation}

The coefficients we have introduced obey the following recursion relations which naturally come from\eqref{eq22}:
\begin{eqnarray}
\begin{split}
\label{eq39}
\tilde{f}_{n+1,0}&=- \tilde{f}_{n,2} - \sqrt{2} \alpha \tilde{f}_{n,1} -  \alpha^2 \tilde{f}_{n,0}; \\
\tilde{f}_{n+1,1}&=-3 \tilde{f}_{n,3} -2 \sqrt{2} \alpha \tilde{f}_{n,2} - (x_1 + \alpha^2) \tilde{f}_{n,1} - \sqrt{2} \alpha x_1\tilde{f}_{n,0} ;\\
\tilde{f}_{n+1,2}&=-6 \tilde{f}_{n,4} -3 \sqrt{2} \alpha \tilde{f}_{n,3} - (2x_1 + \alpha^2) \tilde{f}_{n,2} - \sqrt{2} \alpha x_1 \tilde{f}_{n,1} ;\\
\tilde{f}_{n+1,l}&=-\frac{1}{2} (l+1)(l+2) \tilde{f}_{n,l+2} - \sqrt{2} \alpha (l+1) \tilde{f}_{n,l+1} - (x_1 l + \alpha^2) \tilde{f}_{n,l} - \sqrt{2} \alpha x_1 \tilde{f}_{n,l-1}\\
&+ \frac{1}{3} x_1^2 (l-2)\tilde{f}_{n, l-2}+\frac{\sqrt{2}}{3}\alpha x_1^2 \tilde{f}_{n, l-3}\  (3 < l < 3n-2); \\
\tilde{f}_{n+1,3n-1}&= -3n \sqrt{2} \alpha \tilde{f}_{n,3n} -\big\lbrack(3n-1)x_1+ \alpha^2\big\rbrack \tilde{f}_{n,3n-1} - \sqrt{2} \alpha x_1 \tilde{f}_{n,3n-2}\\
&+ \frac{1}{3} x_1^2 (3n-3)\tilde{f}_{n, 3n-3}+\frac{\sqrt{2}}{3}\alpha x_1^2 \tilde{f}_{n, 3n-4}; \\
\tilde{f}_{n+1,3n}&= -(3n x_1+ \alpha^2) \tilde{f}_{n,3n}- \sqrt{2} \alpha x_1 \tilde{f}_{n,3n-1}+ \frac{1}{3} x_1^2 (3n-2)\tilde{f}_{n, 3n-2}+\frac{\sqrt{2}}{3}\alpha x_1^2 \tilde{f}_{n, 3n-3};\\
\tilde{f}_{n+1,3n+1}&= - \sqrt{2} \alpha x_1 \tilde{f}_{n,3n}+ \frac{1}{3} x_1^2 (3n-1)\tilde{f}_{n, 3n-2}+\frac{\sqrt{2}}{3}\alpha x_1^2 \tilde{f}_{n, 3n-2};\\
\tilde{f}_{n+1,3n+2}&= n x_1^2 \tilde{f}_{n,3n} +\frac{\sqrt{2}}{3}\alpha x_1^2 \tilde{f}_{n, 3n-1};\\
\tilde{f}_{n+1,3n+3}&= \frac{\sqrt{2}}{3}\alpha x_1^2 \tilde{f}_{n, 3n}. 
\end{split}
\end{eqnarray}
We now have to compute the integrals
\begin{equation}
\label{eq38}
C_{m,n}=\sum_{k=0}^{m}\sum_{p=0}^{n} \int_{-\infty}^{+\infty} dq. q^{k+p+1} \tilde{f^\star}_{m,k}(\alpha) \tilde{f}_{n,p}(\alpha) \exp{(2\beta q+x_1 q^2-\frac{1}{6}x_1^2 q^4)}\,
\end{equation}
with
\begin{equation}
\label{eqaaa01}
2 \beta = \sqrt{2}\left(\alpha + \alpha^\star \right).
\end{equation}
We have for example
\begin{equation}
\label{eq40}
C_{0,0}= \int_{-\infty}^{+\infty} dq. q\  \exp{(2\beta q+x_1 q^2-\frac{1}{6}x_1^2 q^4)}\ .
\end{equation}
The integrals in \eqref{eq38} have the generic form
\begin{equation}
\label{eq41}
J_n = \int_{-\infty}^{+\infty} dq. q^n\  \exp{(2\beta q+x_1 q^2-\frac{1}{6}x_1^2 q^4)}\ .
\end{equation}
At this point, there are two ways to tackle the computation. The first insight is that one may need to compute only the integral $J_0$. The second insight is about recursion relations among the  $ J_n $.\\
We begin with the first insight. Its main interest relies in the fact that it leads to analytical expressions. Its main limitation is that it works only for big values of the real part of the 
parameter $\alpha$ describing the coherent state.\\

In short, from \eqref{eq41}, it could be shown that
\begin{equation}
\label{eq43}
J_n = \frac{1}{2^n} \frac{d^n}{d\beta^n} J_0.
\end{equation}
For convenience, we introduce the function $g$ given by
\begin{equation}
\label{eq50}
g(q)=2\beta q+x_1 q^2-\frac{1}{6}x_1^2 q^4.
\end{equation}
Our integral then becomes 
\begin{equation}
\label{eqa07}
J_0  = \int_{-\infty}^{\infty} dq \ \exp{\big\lbrack g(q)\big\rbrack}.
\end{equation}
To evaluate $J_0$, we shall use the saddle point approximation. 
This is due to the fact that the exponential is a very rapidly decaying function, due to the fourth degree term with a negative sign in the exponential. 
The extrema of the integrand satisfy the equation
\begin{equation}
\label{eq44}
q^3-\frac{3}{x_1}q-\frac{3\beta}{x_1^2}=0.
\end{equation}
This equation is a particular case of the following
\begin{equation}
\label{eq45}
q^3+a_1q^2+a_2q+a_3=0.
\end{equation}
The solution to such an equation can be recast using the Cardan formula\cite{Art01-71}. One introduces the intermediate quantities
\begin{equation}
\label{eq46}
Q=\frac{-1}{x_1};\  R=\frac{3\beta}{2x_1^2};
\end{equation}
\begin{equation}
\label{eq47}
D = Q^3+R^2;\ S=\sqrt[3]{R+\sqrt{D}};\ T=\sqrt[3]{R-\sqrt{D}}.
\end{equation}
The only real extremum (actually a maximum) is found at
\begin{equation}
\label{eq48}
q_0= S+T-\frac{1}{3}a_1.
\end{equation}
One finally arrives at the following expression
\begin{equation}
\label{eq49}
q_0=\sqrt[3]{\frac{3\beta}{2x_1^2}+\sqrt{\frac{9\beta^2}{4x_1^4}-\frac{1}{x_1^3}}}+\sqrt[3]{\frac{3\beta}{2x_1^2}-\sqrt{\frac{9\beta^2}{4x_1^4}-\frac{1}{x_1^3}}}.
\end{equation}
Let us first expand the function $g$, given by \eqref{eq50}, near its minimum
\begin{equation}
\label{eqa10}
g(q) = g(q_0) + \frac{1}{2!} g''(q_0) ( q - q_0)^2 + \frac{1}{3!} g'''(q_0) ( q - q_0)^3 + \frac{1}{4!} g^{(4)}(q_0) ( q - q_0)^4 + \cdots
\end{equation}
Introducing the centered and rescaled variable $y$ by
\begin{equation}
\label{eqa11}
q = q_0 + \frac{y}{\sqrt{- g''(q_0)}}
\end{equation}
one obtains
\begin{equation}
\label{eqa14}
g(q) = g(q_0) - \frac{1}{2!} y^2 + A y^3 + B y^4 + \cdots
\end{equation}
where the parameters $A$ and $B$ are given by
\begin{equation}
\label{eqa16a}
A = \frac{1}{3!} \ \frac{g^{(3)}(q_0)}{\big\lbrack - g''(q_0) \big\rbrack^{\frac{3}{2}}}, \ B = \frac{1}{4!} \ \frac{g^{(4)}(q_0)}{\big\lbrack - g''(q_0) \big\rbrack^2}.
\end{equation}
This leads to a sum of gamma functions 
\begin{equation}
\label{eqa17}
J_0 \simeq \frac{\exp\big\lbrack g(q_0)\big\rbrack}{\sqrt{-g''(q_0)}} \ \int_{-\infty}^{+\infty} dy \exp{\left(-\frac{1}{2}y^2\right)} \ \exp{\left(Ay^3+By^4 + \cdots \right)}.
\end{equation}
From this, we shall derive the domain of validity of our approach. To have an asymptotic series for the quantity under investigation, we need $A$ and $B$ to be negligible. A plot of these functions shows this to be true for big values of the parameter $\beta$.

\pagebreak

\begin{figure}[ht]
\centering
\includegraphics[scale=0.64]{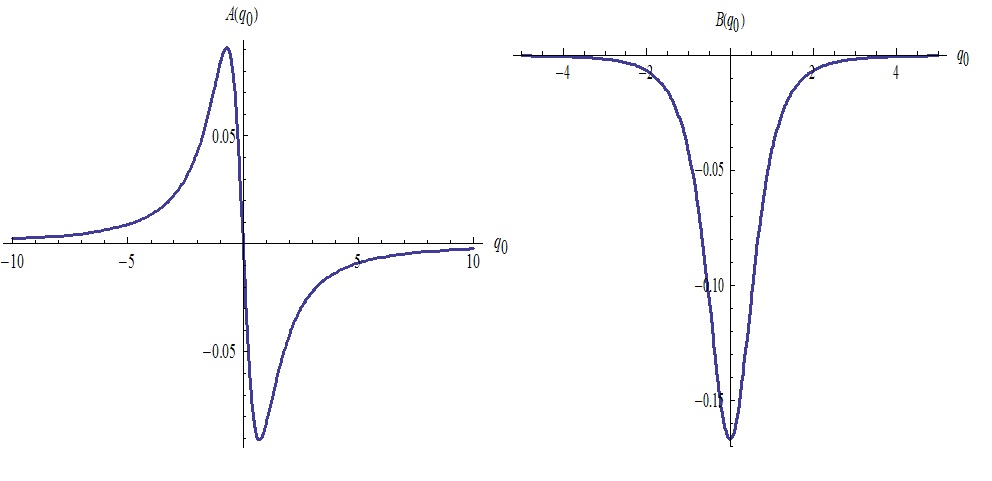}
\caption{{\footnotesize{Plot of the functions $A$ and $B$ for $x_1=-1$.}}}
\end{figure}
We use this to simplify our formulas. For large values of $\beta$, one has that the maximum of the function $g$ occurs at
\begin{equation}
\label{eqa15}
q_0 \simeq \left( \frac{3\beta}{x_1^2} \right)^{\frac{1}{3}},
\end{equation}
(which comes from \eqref{eq49}) because in that limit
\begin{equation}
\label{eqa16b}
A \sim \frac{1}{\beta^{\frac{1}{3}}}, \  B \sim \frac{-4 x_1^2}{\beta^{\frac{4}{3}}}.
\end{equation}
One then finds the dominant contribution (in the limit $\beta \longrightarrow \infty)$ to be given by
\begin{equation}
\label{eqa21}
J_0 \simeq \sqrt{\pi} \exp{\left( z_0 \beta^{\frac{2}{3}}+ z_1  \beta^{\frac{4}{3}} \right)} . \ \left( z_2 + z_3 \beta^{\frac{2}{3}} \right)^{-\frac{1}{2}}
\end{equation}
where the constant coefficients are given by
\begin{equation}
\label{eqa22}
z_0 = -x_1 \left( \frac{3\beta}{x_1^2} \right)^{\frac{2}{3}}, \  z_1 = \frac{x_1^2}{2} \left( \frac{3\beta}{x_1^2} \right)^{\frac{4}{3}}, \  z_2 = -x_1, \  z_3 = x_1^2 \left( \frac{3\beta}{x_1^2} \right)^{\frac{2}{3}}.
\end{equation}
For the rest of our treatment, it appears more simple to write this result directly in terms of $q_0$ :
\begin{equation}
\label{eqa23}
J_0 = \sqrt{\pi} \left( -x_1 + x_1^2 q_0^2 \right)^{-\frac{1}{2}}\exp{(-x_1 q_0^2 + \frac{1}{2} x_1^2 q_0^4)}.
\end{equation}

The second insight uses integration by parts. One has
\begin{equation}
\label{eqa25a}
J_n = \int_{-\infty}^{+\infty} dq \ q^n \ \exp{\big\lbrack g(q) \big\rbrack}.
\end{equation}              
If one introduces $u = q^n \ \exp{(x_1 q^2 - \frac{1}{6} x_1^2 q^4)}$ and $dv = \exp{(2\beta \ q)}dq$, one gets
\begin{eqnarray}
\begin{split}
\label{eqa26}
J_3 &= \frac{3}{x_1} J_1 + \frac{3\beta}{x_1^2} J_0 , \\
J_{n+3} &= \frac{3n}{2 \ x_1^2} J_{n-1}+ \frac{3\beta}{x_1^2} J_n + \frac{3}{x_1} J_{n+1} \ ; n \geq 1.
\end{split}
\end{eqnarray}
These relations are exact. They apply to small as well as to large values of $\beta$. They could be used in the following way. For a given value of $\beta$, one computes numerically $J_0, J_1$ and $J_2$. All the others are then obtained by the preceding recursion formulas.\\
One can now use the relation between $J_n$ and $J_0$. For example, one gets
\begin{eqnarray}
\begin{split}
\label{eqa25}
J_1 = \frac{\sqrt{\pi}}{2} q_0^{-1}& \left( -x_1 + x_1^2 q_0^2 \right)^{-\frac{3}{2}} \left( 1 - 4 x_1 q_0^2 + 2 x_1^2 q_0^4 \right) \exp{(-x_1 q_0^2 + \frac{1}{2} x_1^2 q_0^4)}, \\
J_2 = \frac{\sqrt{\pi}}{4} x_1^{-1} q_0^{-4}& \left( -x_1 + x_1^2 q_0^2 \right)^{-\frac{5}{2}}\\
&\times ( 1 +2 x_1 q_0^2 -10 x_1^2 q_0^4 + 22 x_1^3 q_0^6 -16 x_1^4 q_0^8 + 4 x_1^5 q_0^{10} ) \exp{(-x_1 q_0^2 + \frac{1}{2} x_1^2 q_0^4)}.
\end{split}
\end{eqnarray}
It should be noted that only the dominant term in the polynomial expression is relevant. This is justified by the fact that the contribution coming from the coefficients $A$ and $B$ in \eqref{eqa17} are ignored when computing the dominant term. This approach, which gives analytical formulas, will be used later in the computation of the time scale where the classical solution begins separating from the mean value of the position operator for the coherent states studied here.\\

Finally, we use the mean value of the position operator given in \eqref{eq28}. We first write explicitly the relation contained in \eqref{eq29}
\begin{equation}
\label{eqa03}
C_{m,n} = \sum _{k=0}^{3m} \sum _{p=0}^{3n} \tilde{f}^\star_{m,k} \ \tilde{f}_{n,p} \ J_{k+p+1}.
\end{equation}  
The expressions of the $C_{m, n}$ (in \eqref{eqa03}) then leads to
\begin{eqnarray}
\begin{split}
\label{eqa04}
\Omega_0 &= J_1,  \\
\Omega_1 = i \ & \Big\lbrack  \left( - \tilde{f}_{1,0} + \tilde{f}^\star_{1,0} \right) J_1 + \left( - \tilde{f}_{1,1} + \tilde{f}^\star_{1,1} \right) J_2 \Big\rbrack, \\
\Omega_2 = - \frac{1}{2} & \ \Big\lbrace \  \left( \tilde{f}_{2,0} + \tilde{f}^\star_{2,0} - 2\tilde{f}^\star_{1,0} \tilde{f}_{1,0} \right) J_1 \\
&+ \Big\lbrack \tilde{f}_{2,1} + \tilde{f}^\star_{2,1} - 2 \left( \tilde{f}^\star_{1,0} \tilde{f}_{1,1} + \tilde{f}_{1,0} \tilde{f}^\star_{1,1} \right) \Big\rbrack J_2 + \ \left( \tilde{f}_{2,2} + \tilde{f}^\star_{2,2} - 2\tilde{f}^\star_{1,1} \tilde{f}_{1,1} \right ) J_3 \ \Big\rbrace, \\
\Omega_3 =  \frac{i}{6} & \ \Big\lbrace \ \Big\lbrack \left( \tilde{f}_{3,0} - \tilde{f}^\star_{3,0} \right) + 3 \left( - \tilde{f}^\star_{1,0} \tilde{f}_{2,0} + \tilde{f}_{1,0} \tilde{f}^\star_{2,0} \right) \Big\rbrack J_1 \\
&+ \Big\lbrack \left( \tilde{f}_{3,1} - \tilde{f}^\star_{3,1}\right) + 3 \left( -\tilde{f}^\star_{1,0} \tilde{f}_{2,1} - \tilde{f}_{2,0} \tilde{f}^\star_{1,1} + \tilde{f}_{1,0}\tilde{f}^\star_{2,1} + \tilde{f}_{1,1} \tilde{f}^\star_{2,0} \right) \Big\rbrack J_2 \\ 
&+ \ \Big\lbrack \left( \tilde{f}_{3,2} - \tilde{f}^\star_{3,2} \right) + 3 \left( -\tilde{f}^\star_{1,0} \tilde{f}_{2,2} - \tilde{f}_{2,1} \tilde{f}^\star_{1,1} + \tilde{f}_{1,0}\tilde{f}^\star_{2,2} + \tilde{f}_{1,1} \tilde{f}^\star_{2,1} \right) \Big\rbrack J_3 \\
& + \Big\lbrack \left( \tilde{f}_{3,3} - \tilde{f}^\star_{3,3} \right) + 3 \left( - \tilde{f}^\star_{1,1} \tilde{f}_{2,2} + \tilde{f}_{1,1} \tilde{f}^\star_{2,2} \right)  \Big\rbrack J_4 \Big\rbrace,\\
&\cdots
\end{split}
\end{eqnarray}
On the other hand, the classical trajectories are analytical functions of time. Rather than relying on their periodic character, we can also consider the equation of motion
\begin{equation}
\label{eq64}
q_{cl}''(t)-\mu q_{cl}^3(t)-\sigma q_{cl}^5(t)=0,
\end{equation}
\hspace{40mm}with $ \mu=-\frac{4}{3} \omega^2 x_1^3$ and $ \sigma= \frac{1}{3} \omega^2 x_1^4 $.\\
One could see that a power series solution of the form
\begin{equation}
\label{eq65}
q_{cl}(t)=\sum_{i=0}^{\infty} A_i . (\omega t)^i
\end{equation}
leads to the following relations between the coefficients
\begin{eqnarray}
\begin{split}
\label{eq66}
A_2 & = \frac{1}{6}(A^5_0 - 4A^3_0); \\  
A_3 & = \frac{1}{18}( 5A_1 A^4_0 - 12A_1 A^2_0); \\ 
A_4 & = \frac{1}{216}( 5A_0^9 - 32A^7_0 + 48A^5_0 + 60A^2_1 A^3_0 - 72 A^2_1 A_0); \\ 
A_5 & = \frac{1}{1080}(85A_1 A^8_0 - 432A_1 A^6_0 + 432 A_1 A^4_0 + 180 A^3_1 A^2_0 - 72 A^3_1);\\
&\cdots
\end{split}
\end{eqnarray}
The classical solution of the system and the mean value of the position operator will be identical, if the equations in \eqref{eq66} are still satisfied when the constants $A_i$ are replaced by the coefficients $\Omega_i$. For our case, this amounts to the vanishing of the quantity $q_{\text{cl}}(t)-q_{\text{moy}}(t)$. For the harmonic oscillator, this quantity exactly vanishes for the coherent states defined by Schr\"{o}dinger\cite{Art01-05}. For systems where the classical and "quantum" trajectories do not coincide, one could introduce the quantity:
\begin{equation}
\label{eqz01}
\varepsilon(t)=\frac{q_{cl}(t) - q_{moy} (t)}{q_{cl} (t)}.
\end{equation}
When $\lvert \varepsilon(t)\rvert \ll 1$, the classical and "quantum" trajectories are close. This means 
\begin{equation}
\label{eqz02}
\lvert \omega t \rvert \leq \left( \varepsilon(t) \ \Big \lvert \frac{x_1^2}{f(x_1, \beta)} \Big \rvert\right)^{\frac{1}{2}},
\end{equation}
where
\begin{eqnarray}
\begin{split}
\label{eqz03}
f(x_1, \beta) = &6x_1 \Big\lbrack -4 \gamma^2 \left( 3x_1 \beta^4 \right)^{\frac{1}{3}} - \left( 9 x_1^2 \beta^2 \right)^{\frac{1}{3}} \left( \beta^2 + \gamma^2 \right) - 2 \beta^2 \gamma^2 + x_1 \left( \beta^2 - \gamma^2 \right) \Big\rbrack\\
& + \pi^2 \exp{\bigg\lbrack 4 \left( -x_1 q_0^2 + \frac{1}{2} x_1^2 q_0^4 \right) \bigg\rbrack} - 4 \pi x_1^2 \exp{\bigg\lbrack 2 \left( -x_1 q_0^2 + \frac{1}{2} x_1^2 q_0^4 \right) \bigg\rbrack}.
\end{split}
\end{eqnarray}
This leads to the following relation when an initial time  $t_o$ has been chosen and  $\varepsilon(t_o) $  replaced by  $\varepsilon $:
\begin{equation}
\label{eqz04}
\lvert \omega t \rvert \leq \left( \varepsilon \ \Big \lvert \frac{6 \Omega_0}{ \Omega_0^5  - 4 \Omega_0^3 - 6 \Omega_2} \Big \rvert \right)^{\frac{1}{2}} 
\end{equation}
This will hold true for times very small compared to the intrinsic frequency of the system.\\

Some mathematical issues have not been addressed here. One of them is the convergence of the series obtained for the mean value of the position operator. This is a very difficult point since we do not have an explicit, simple formula for the coefficient $\Omega_l$ for all $l$. But in principle, since the problem is well posed,  the results should be meaningful.\\ 
The method we followed here gives the position as an analytic series in time. One has to sum a finite number of terms to obtain an approximation. Such a finite sum is polynomial. This explains why after some time the graph goes to infinity; this simply means we are out of the domain where the approximation is valid. The SUSY structure has been extensively used (see \eqref{eq17}, \eqref{eq18}, \eqref{eq19}). This culminated in the recurrence formula  \eqref{eq22} which gives the $n^{th}$ contribution to the wave function. The superpotential $x(q)$ appears many times because it is the result of the commutation relation between the operators $\hat{A}$ and $\hat{A}^\dag$ (See \eqref{eq03a}).\\
The coefficients $C_{m,n}$ are not easy to find analytically, even for simple superpotentials like the one we studied here. We devised recipes which can tackle this successfully. This was the case with the saddle point approximation whose domain of validity was explicitly given.\\

Let us analyse  some considerations concerning formal remarks which can be misleading. It has been argued, in the case of coherent states defined as infinite superpositions of energy eigenstates, that in most cases the wave function so obtained is only formal i.e it does not converge when considering for example imaginary times. We have no reason to turn to imaginary times in this work.\\

\section{Conclusions}
\

The results of our paper can be summarized as follows\\

For the harmonic oscillator, the approach we propose successfully reproduces what has been found by other methods\cite{Art01-05,Art01-16,Art01-19}.\\

We then analysed two other specific models. The first one, which is the core of the article, is described by a third degree superpotential. The computation of the mean value of the position operator for the coherent states defined by Molski\cite{Art01-01} was performed. The second one was a family of superpotentials which are isospectral to the harmonic oscillator. We expect other workers in the field will analyze these systems and compare their results to ours, using different methods.\\

Concerning the perspectives, we wish to study systems such as the Poschl-Teller and the Morse potentials among others. It is likely that numerical treatment will be more important than in the cases treated here.

\pagebreak

\appendix

\section{The Harmonic Oscillator}
\

We give here the results obtained by our treatment when applied to the harmonic oscillator. The functions $f_n$ are polynomial
\begin{equation}
\label{eq69}
f_n(q, \alpha)=\sum_{k=0}^n \tilde{f}_{n,k}(\alpha) q^k; \tilde{f}_{0,0}= \tilde{f}_{0,0}^\star=1.
\end{equation}
The coefficients we need are given by the following integrals
\begin{eqnarray}
\begin{split}
\label{eq70}
C_{m,n}&=\sum_{k=0}^{m}\sum_{p=0}^{n} \int_{-\infty}^{+\infty} dq. q^{k+p+1} \tilde{f}^\star_{m,k}(\alpha) \tilde{f}_{n,p}(\alpha) \psi_H^\star (q,\alpha) \psi_H (q,\alpha)\ ; \\ 
       &=\sum_{k=0}^{m}\sum_{p=0}^{n} \int_{-\infty}^{+\infty} dq. q^{k+p+1} \tilde{f^\star}_{m,k}(\alpha) \tilde{f}_{n,p}(\alpha) \exp{(2\beta q-q^2)}\ 
\end{split}
\end{eqnarray}
where, by definition, the coefficient $\beta$ is given by \eqref{eqaaa01}.\\

We have fewer recursion relations
\begin{eqnarray}
\begin{split}
\label{eq72}
\tilde{f}_{n+1,0}&=- \tilde{f}_{n,2} - \sqrt{2} \alpha \tilde{f}_{n,1} - \alpha^2 \tilde{f}_{n,0}; \\
\tilde{f}_{n+1,l}&=-\frac{1}{2} (l+1)(l+2) \tilde{f}_{n,l+2} - \sqrt{2} \alpha (l+1) \tilde{f}_{n,l+1} + ( l - \alpha^2) \tilde{f}_{n,l} + \sqrt{2} \alpha \tilde{f}_{n,l-1};\\
\tilde{f}_{n+1,n-1}&=- \sqrt{2} \alpha n \tilde{f}_{n,n} +(n-1-\alpha^2) \tilde{f}_{n,n-1} + \sqrt{2} \alpha \tilde{f}_{n,n-2}; \\ 
\tilde{f}_{n+1,n}&=(n-\alpha^2)\tilde{f}_{n,n} + \sqrt{2} \alpha \tilde{f}_{n,n-1}; \\
\tilde{f}_{n+1,n+1}&= \sqrt{2} \alpha \tilde{f}_{n,n}.
\end{split}
\end{eqnarray}
Our coefficients $C_{m,n}$ are given by integrals of the product of an exponential and a power of the variable $q$ (See \eqref{eq70}). Actually, the only thing one could need is
\begin{equation}
\label{eq73}
J_0 =\int_{-\infty}^{+\infty} dq. \exp{(-q^2+2\beta q)}\,
\end{equation}
because
\begin{equation}
\label{eq74}
\int_{-\infty}^{+\infty} dq. q^{k+p+1}\exp{(2\beta q-q^2)}\ =\sqrt{\pi}\  \frac{1}{2^{k+p+1}}\  \frac{\partial^{k+p+1}}{\partial \beta^{k+p+1}} \Big\lbrack \exp{(\beta^2)} \Big\rbrack.
\end{equation}
The coefficients can now be recast in the form
\begin{equation}
\label{eq75}
C_{m,n}=\sqrt{\pi} \exp{\left(\beta^2\right)}\sum_{k=0}^{m}\sum_{p=0}^{n} \tilde{f^\star}_{m,k}(\alpha)f_{n,p}(\alpha)\  \frac{1}{2^{k+p+1}}\   P_{k+p+1}(\beta).
\end{equation}
The polynomials $P_s$ are defined by the property
\begin{equation}
\label{eq76}
\frac{d^s}{d\beta^s}\exp{(\beta^2)}=\exp{(\beta^2)}\  P_s(\beta).
\end{equation}
One readily finds that they obey the recursion relations
\begin{eqnarray}
\label{eq77}
P_{s+1}(\beta)&=&2\beta P_s(\beta) + \frac{d}{d\beta}P_s(\beta); \\  \nonumber
P_0(\beta)&=&1.
\end{eqnarray}
Let us now compare the mean value of the position operator and the classical trajectory for the harmonic oscillator using our approach.\\
For the quantum behavior, one finds
\begin{eqnarray}
\begin{split}
\label{eq80}
\Omega_0 = \sqrt{\pi} \ \beta \ \exp{(\beta^2)}, & \  \Omega_1 = i \sqrt{\frac{\pi}{2}} \ \beta \exp{(\beta^2)},\\
\Omega_2 =-\frac{1}{2} \sqrt{\pi}\ \beta \exp{(\beta^2)}, & \  \Omega_3 =-\frac{i}{6} \sqrt{\frac{\pi}{2}} \ \beta  \ \exp{(\beta^2)},\\
\Omega_4 =\frac{1}{24} \sqrt{\pi}\ \beta \exp{(\beta^2)}, & \  \cdots
\end{split}
\end{eqnarray}
The following ratios
\begin{eqnarray}
\begin{split}
\label{eq85}
&\frac{\Omega_2}{\Omega_0}=-\frac{1}{2},\ \frac{\Omega_4}{\Omega_0}=\frac{1}{24},\ \frac{\Omega_6}{\Omega_0}=-\frac{1}{720},\ \frac{\Omega_8}{\Omega_0}=\frac{1}{40 320},\ \frac{\Omega_{10}}{\Omega_0}=-\frac{1}{3 628 800},\ \cdots \\
&\frac{\Omega_3}{\Omega_1}=-\frac{1}{6},\ \frac{\Omega_5}{\Omega_1}=\frac{1}{120},\ \frac{\Omega_7}{\Omega_1}=-\frac{1}{5 040},\ \frac{\Omega_9}{\Omega_1}=\frac{1}{362 880},\ \cdots
\end{split}
\end{eqnarray}
drives us to conclude that the behavior of the classical trajectory given in \eqref{eqq01} is recovered, at least in the lowest orders. One can go to higher order and verify this still works.\\
\begin{figure}[ht]
\centering
\includegraphics[scale=0.60]{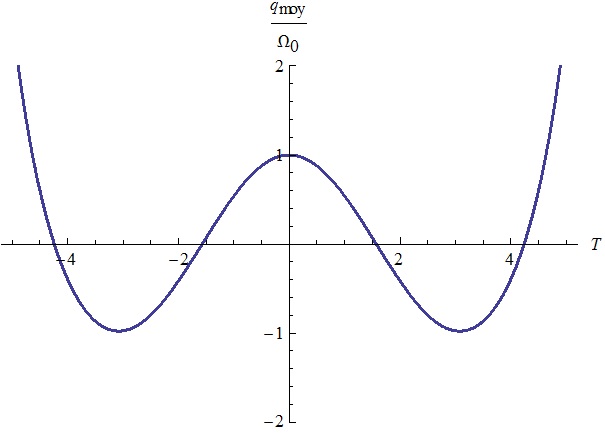}
\caption{{\footnotesize{Plot of $q_{moy}$ for the Harmonic Oscillator, for the value $\alpha=10$. $8^{th}$ Order Approximation.}}}
\end{figure}

The same calculation done in section 4, for our toy model, shows the plot of $q_{moy}(t)$ below.
\begin{figure}[ht]
\centering
\includegraphics[scale=0.60]{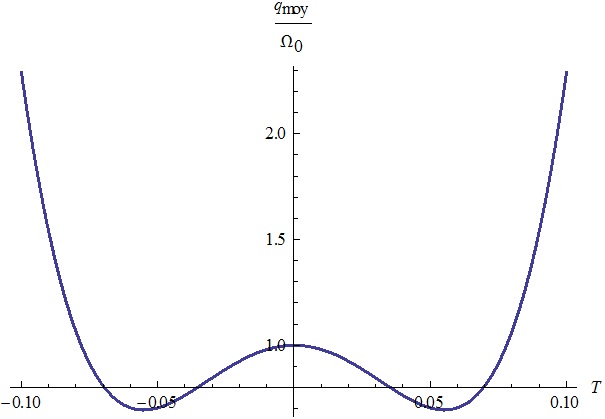}
\caption{{\footnotesize{Plot of $q_{moy}$ for the Toy Model for the values $\alpha=\frac{100\sqrt{2}}{3}-\frac{i}{10000}$ and $x_1=-1$.The calculation was done to the $4^{th}$ order.}}}
\end{figure}

\pagebreak

\section{A Family of Superpotentials encompassing The Harmonic Oscillator}
\

The superpotential of a system is not unique. For systems who have Hamiltonians isospectral to the harmonic oscillator one, we have the family of superpotentials $x_{\lambda}$ given by \cite{Art01-67}:
\begin{equation}
\label{eq90}
x_{\lambda}(q) = q + \frac{\exp{(-q^2)}}{\lambda + \frac{\sqrt{\pi}}{2} \ \mathrm{erf(q)}},
\end{equation}
\hspace{50mm}where $\mathrm{erf}$ is the error function.\\
Notice that the convention used here is different from the one used by \cite{Art01-01} but leads to the same physics.\\
The general superpotential given by \eqref{eq90} is obtained by the resolution of a Riccati equation.\\
The wave function of a coherent state is given by
\begin{equation}
\label{eq91}
\psi_H (q) =  N \exp{\Big\lbrack \sqrt{2} \alpha q - \int_0^q d\xi \ x_{\lambda} (\xi) \Big\rbrack}.
\end{equation}
After the calculation of the integral in \eqref{eq91} and using the fact that 
\begin{equation}
\label{eq92}
\frac{d}{dq} \Big\lbrack \mathrm{erf}(q) \Big\rbrack = \frac{2}{\sqrt{\pi}} \exp{(-q^2)},
\end{equation}
one finds
\begin{equation}
\label{eq93}
\psi_H (q) =  N \left( 1 + \frac{\sqrt{\pi}}{2 \lambda} \mathrm{erf}(q) \right)^{-1} \exp{\left( \sqrt{2} \alpha q - \frac{q^2}{2}\right)}.
\end{equation}
A plot of the mean value of the position operator is given in Fig.5.
\begin{figure}[ht]
\centering
\includegraphics[scale=0.70]{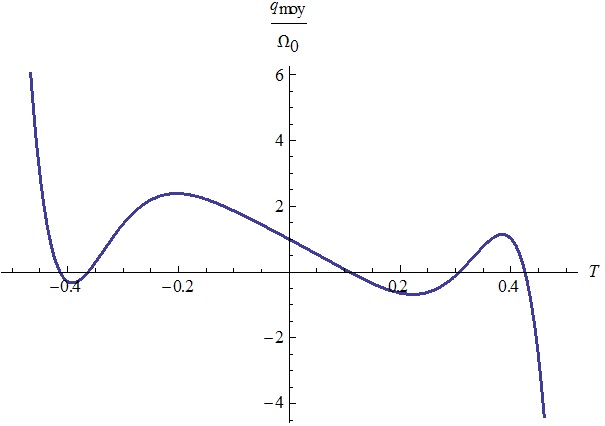}
\caption{{\footnotesize{Plot of $q_{moy}$ for the values $\alpha= 1+i$, $\lambda = 10^9$. The calculation was done to the $10^{th}$ order.}}}
\end{figure}

\pagebreak

\end{document}